\documentclass[twocolumn]{aastex62}

\usepackage{lineno}
\shorttitle{Optical/$\gamma$-ray Blazar Flare Correlations}
\shortauthors{Liodakis et al.}

\begin{document}

\title{Probing Blazar Emission Processes with Optical/Gamma-ray Flare Correlations}


\correspondingauthor{I. Liodakis}
\email{ilioda@stanford.edu}

\author{Ioannis Liodakis}
\affil{KIPAC, Stanford University, 452 Lomita Mall, Stanford, CA 94305, USA}

\author{Roger W. Romani}
\affil{KIPAC, Stanford University, 452 Lomita Mall, Stanford, CA 94305, USA}

\author{Alexei V. Filippenko}
\affil{Department of Astronomy, University of California, Berkeley, CA 94720-3411, USA}
\affil{Miller Senior Fellow, Miller Institute for Basic Research in Science, University of California, Berkeley, CA  94720, USA}

\author{Daniel Kocevski}
\affil{Astrophysics Office, ST12, NASA/Marshall Space Flight Center, Huntsville, AL 35812, USA}

\author{WeiKang Zheng}
\affil{Department of Astronomy, University of California, Berkeley, CA 94720-3411, USA}

\begin{abstract}
Even with several thousand {\it Fermi}-LAT blazar detections, the $\gamma$-ray emission mechanism is 
poorly understood. We explore correlated optical/$\gamma$-ray flux variations for 178 {\it Fermi}-LAT blazars regularly monitored by KAIT, SMARTS, and the Steward Observatory. Out of the 178 sources, 
121 show a measurable ($>1\sigma$) discrete correlation function peak. Using the derived time-lags and Bayesian block
light-curve decompositions, we measure the fraction of common and orphan flares between the two bands. 
After accounting for sampling and sensitivity limitations we quantify for the first time the true orphan flare rates of optical and $\gamma$-ray flares: 54.5\% of optical and 20\% of $\gamma$-ray 
flares are orphan events. Both the intraband temporal relation and the small orphan $\gamma$-ray flare  fraction point toward leptonic processes as the likely mechanism for the high-energy emission. Motivated to discriminate between synchrotron self-Compton and external-Compton dominance in individual sources, we use the flux-flux variations to determine the slope $m$ of the 
${\rm log}\,f_{\rm opt} - {\rm log}\,f_{\gamma}$ dependence. The slope distribution
suggests a bimodal population with high and intermediate synchrotron-peak objects showing larger $m$ than 
low synchrotron-peak objects.  We find that $m$ is naturally decreased through pollution from the orphan 
(typically optical) flares and develop a method to statistically recover, given the sources' measured orphan flare
rate, the intrinsic $m$. While source classes show composite behavior, the majority of BL Lac objects favor $m=2$, indicating a synchrotron self-Compton origin
for the $\gamma$-rays. No preference for either $m$ is found in flat spectrum radio quasars.
\end{abstract}

\section{Introduction}\label{sec:intro}

Blazars dominate the extragalactic $\gamma$-ray sky, accounting for more than 98\% of the active 
galactic nuclei (AGNs) detected by the Large Area Telescope (LAT) onboard the  {\it Fermi} $\gamma$-ray space observatory \citep{Acero2015}. 
This is largely caused by relativistic beaming of emission from jets directed toward our line of sight 
\citep{Blandford1979}. Blazar spectral energy distributions (SEDs) have a characteristic two-hump shape with their typically highly variable emission covering the entire electromagnetic spectrum (e.g., \citealp{Urry1996,Fossati2000,Marscher2008,Abdo2010-III,Liodakis2017-IV,Liodakis2018-II}). Sources in the blazar population are traditionally classified according to presence
(flat spectrum radio quasars; FSRQs) or absence (BL Lac objects; BL Lacs) of strong, broad
resonance lines in their optical spectra. A different classification scheme uses the frequency
of the synchrotron peak ($\nu_{\rm sy}$) to identify low synchrotron peak (LSP), intermediate synchrotron peak (ISP), 
and high synchrotron peak (HSP) sources, with $\nu_{\rm sy}$ in the infrared, optical, or ultraviolet/X-ray 
bands, respectively.

While the low-energy component of the SED is well understood to be 
synchrotron emission of relativistic electrons spiraling in the jet magnetic field (e.g., \citealp{Raiteri2017}), the nature of the high-energy SED component is still a matter of debate. Perhaps the most likely process
is inverse-Compton (IC) scattering of infrared--to--X-ray photons, with this seed flux being
either jet photons (synchrotron self-Compton; SSC; e.g., \citealp{Marscher1985}) or thermal photons from surrounding structures
such as the broad-line regions, the disk, or the torus (external-Compton; EC; e.g., \citealp{Dermer1992}). Other 
possible mechanisms include hadronic processes (photo-pion production, proto-synchrotron, etc.; \citealp{Mannheim1993}).  Typically one addresses this question
via detailed SED modeling of data collected from intensive multiwavelength campaigns focusing on single 
sources, but such studies are sensitive to the details of particular
outbursts and have not yet proved definitive  (e.g., \citealp{Rani2013,Casadio2015,An2018}).

Another approach is the statistical study of the sources' variability. Long-term correlated variability between the low-energy bands and $\gamma$-rays is generally expected in the leptonic scenario (e.g., \citealp{Zhang2018}) but is not required in the hadronic model. While the SEDs of blazars can be modeled equally well by both leptonic and hadronic processes (e.g., \citealp{Boettcher2012}), the presence (or not) of $\gamma$-ray flares without a low-energy counterpart, often referred to as ``orphan'' flares, could break the degeneracy and identify the dominant process operating in blazar jets. In the case of leptonic emission, flux variations in different bands can also be used to probe the mechanism responsible for the high-energy emission (SSC or EC), since the ratio of synchrotron to inverse-Compton emission is expected to be different for different mechanisms (e.g., \citealp{Bonnoli2011}).

\citet{Liodakis2018} explored the intraband temporal correlations of blazars in radio, optical, and  $\gamma$-rays. Out of all pairs of wavelengths we found the strongest relation between optical and $\gamma$-rays, with most sources showing zero time-lags within our $\sim30$~d resolution. That strong correlation motivates us to further explore this connection. In this paper we focus on the optical--$\gamma$-ray connection in an attempt understand the blazar flaring properties, probe the statistics of the orphan flares, and constrain the source of the target photon fields. Section \ref{sec:samp} presents the sample and the steps of data reduction, and 
in Section \ref{sec:cross} we use the discrete correlation function to quantify intraband temporal correlations. We explore in Section \ref{sec:mult_flares} the correspondence of 
flares between the two bands, and in Section \ref{sec:flare_cor} we measure the scaling between linked optical 
and $\gamma$-ray amplitude variations. Our findings are summarized in Section \ref{sec:disc-conc}.

\section{Sample}\label{sec:samp}

We begin our sample selection with the sources observed by the Katzman Automatic Imaging Telescope \citep[KAIT;][]{Li2003}. KAIT has been observing a total of 152 sources (white-light observations roughly corresponding to the $R$ band) that were detected at high significance ($>10\sigma$) by {\it Fermi}-LAT during the first year of operations \citep{Abdo2010-V} since 2008. To increase our sample and improve our statistics we included sources observed by the Steward Observatory (19 sources) and the Small and Moderate Aperture Research Telescope System (SMARTS, 31 sources) monitoring programs. While these programs include a few additional sources, we restricted our study to sources with substantial (on average $<10$-d cadence) $R$-band/white-light optical coverage during at least 3 consecutive years. The optical observations cover the time period between 2008 and 2017. 

Our final sample consists of 178 $\gamma$-ray-detected sources (80 LSPs, 27 ISPs, 35 HSPs, 36 sources without SED 
peak information)\footnote{The SED classifications of the sources are taken from \citet{Nieppola2006,Nieppola2008,Abdo2010-II,Cohen2014,Lister2015,Mingaliev2015,Angelakis2016}, and references therein.}. Based on optical spectra, 107 sources have been classified as BL Lacs, 64 as FSRQs, 4 as radio galaxies, and 3 are as yet unclassified\footnote{The optical classifications of the sources are taken from SIMBAD \citep{Wenger2000} and the NASA/IPAC Extragalactic Database (NED); NED is operated by the Jet Propulsion Laboratory, California Institute of Technology, under contract with the National Aeronautics and Space Administration (NASA).} covering a redshift range of [0.00021, 2.19]. For the sources in more than one monitoring program, we have combined the observations to increase the sampling density. All apparent magnitudes are converted to $\rm mJy$ using the standard conversion factors for the $R$-band filter \citep{Johnson1966}. Details on the optical data reduction are provided by \citet{Filippenko2001} and \citet{Li2003} for KAIT\footnote{http://herculesii.astro.berkeley.edu/kait/agn/}, \citet{Bonning2012} for 
SMARTS\footnote{http://www.astro.yale.edu/smarts/glast/home.php}, and \citet{Smith2009} for the Steward 
Observatory monitoring program\footnote{http://james.as.arizona.edu/$\sim$psmith/Fermi/}. All photometry is available online from the respective projects' websites.

For the $\gamma$-ray observations we used data from the LAT. We generated $\gamma$-ray light curves through a maximum likelihood optimization technique using the standard analysis tools developed by the LAT collaboration (ScienceTools version v10r01p0)\footnote{http://fermi.gsfc.nasa.gov/ssc/}. In standard unbinned likelihood analysis, the observed distribution of counts at a particular position in the sky is fit to a model that includes all known $\gamma$-ray sources in the 3FGL catalog \citep{Acero2015}, as well as Galactic and isotropic background components\footnote{http://fermi.gsfc.nasa.gov/ssc/data/access/lat/Background Models.html}. The Galactic component (\emph{gll\_iem\_v06}) is a spatial and spectral template that accounts for interstellar diffuse $\gamma$-ray emission from the Milky Way \citep{Fermi_diffuse_model2016}.  The isotropic component (\emph{iso\_source\_v06}) provides a spectral template to account for all remaining isotropic emission including contributions from both residual charged particle backgrounds and the isotropic celestial $\gamma$-ray emission. 

In particular, we used the ``P8R2\_SOURCE\_V6'' instrument response functions and selected `SOURCE' class events in the 0.1--100 GeV energy range from a $12^\circ$ radius region of interest (ROI) centered on the source location with $105^\circ$ as the zenith cut\footnote{While a zenith cut of $90^\circ$ is recommended, it is not unusual for variability studies to adopt a smaller ROI and higher zenith cuts, as both act to reduce the loss of exposure due to the ROI exceeding the zenith cut and overlapping Earth's limb. Such losses could be more significant for the smaller time bins (3~d) as opposed to the larger ones (30~d). We nevertheless use the same ROI and zenith cut regardless of binning, for consistency.}. The flux determination and spectral modeling for each target source is accomplished by fitting the LAT data to the model of the $\gamma$-ray sky, where the normalization and spectral parameters for the target source are left free to vary during the model optimization. The spectral types used for all sources in our fits are drawn from their best-fit 3FGL models. Following the standard procedure, the flux normalization and spectral index were fixed for sources outside the $10^\circ$ radius.  Although the data were fit within an ROI of $12^\circ$, the xml model used in the fit includes sources within ROI $+10^\circ$, so the fits account for contributions from sources outside the ROI. A likelihood-ratio test is then employed to quantify whether there exists a significant excess of counts due to the target source above the expected background model.  If the source is not detected with at least a 3$\sigma$ significance, we calculate the 95$\%$ confidence level upper limits using the Bayesian method described by \cite{Ackermann2016}. Using this procedure we generate 3~d, 10~d, and 30~d $\gamma$-ray light curves. The time period chosen for the analysis covers a mission elapsed time (MET) range of 239,557,418 to 543,167,018, or 2008-08-04 15:43:37 to 2018-03-19 15:43:33 UTC. Throughout the paper, using the spectral index in individual bins, we converted the LAT photon flux (photons\,cm$^{-2}$\,s$^{-1}$) for each source to $\rm mJy$ at 1~GeV (e.g., \citealp{Singal2014}) to match the units of the optical observations.

\section{Cross-correlations}\label{sec:cross}
 \begin{figure*}
\resizebox{\hsize}{!}{\includegraphics[scale=1]{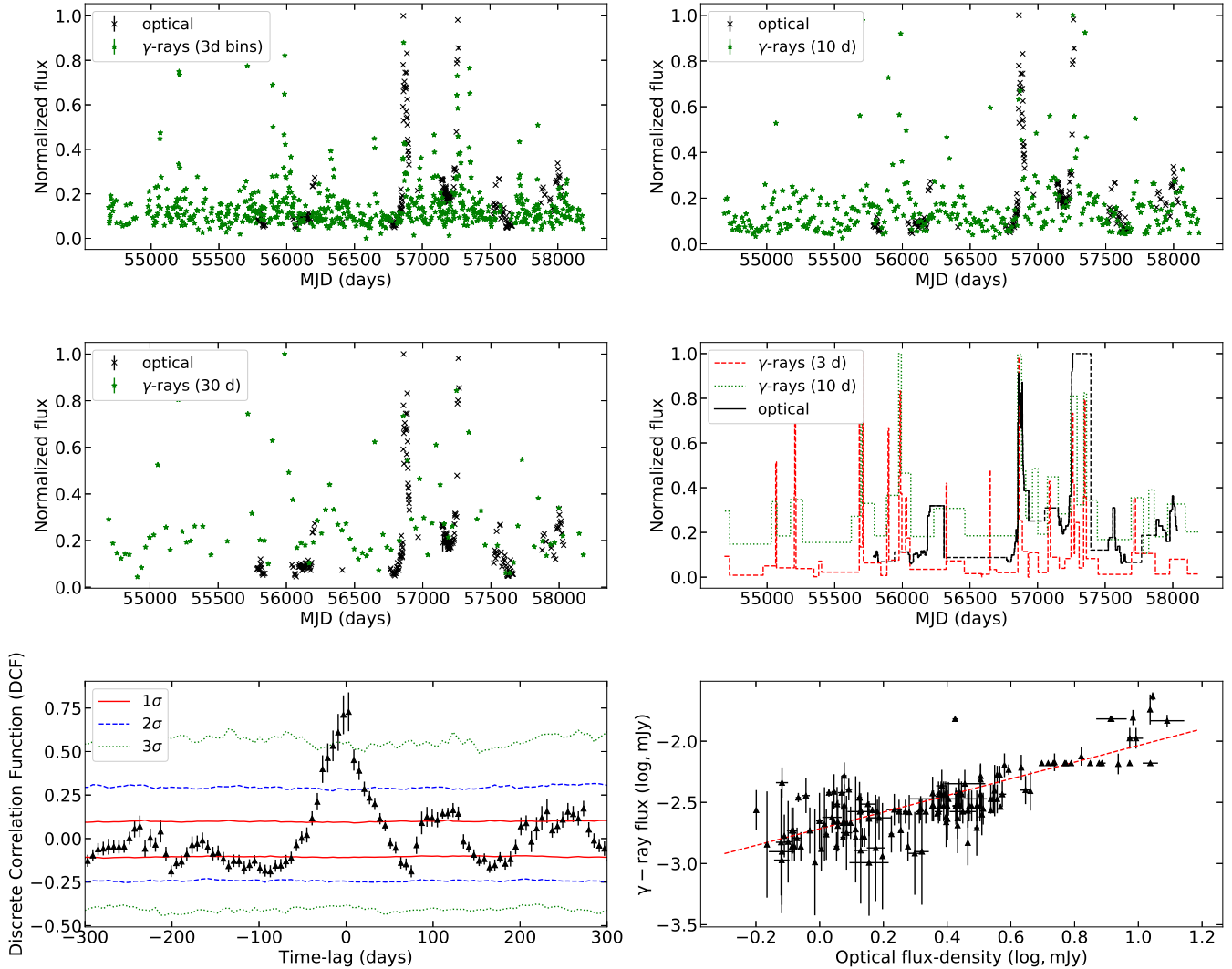} }
 \caption{The top two rows show the optical versus the 3~d, 10~d, and 30~d binned $\gamma$-ray light curves, 
and the Bayesian block representation of the optical and the 3~d and 10~d light curves, for J1800+7828. The fluxes have been normalized to the highest value in each band. The dashed section in the optical blocks 
correspond to seasonal observing gaps. The $\gamma$-ray upper limits have been omitted. The lower two panels show the discrete correlation function (DCF, left) 
and the contemporaneous optical versus $\gamma$-ray fluxes in log--log space. Both optical and $\gamma$-ray fluxes are in mJy. The red dashed line shows the best-fit power-law relation ($Y = (0.69\pm0.05)X + (-2.71\pm0.03)$).}
\label{plt:multi}
\end{figure*}

\begin{figure}
\resizebox{\hsize}{!}{\includegraphics[scale=1]{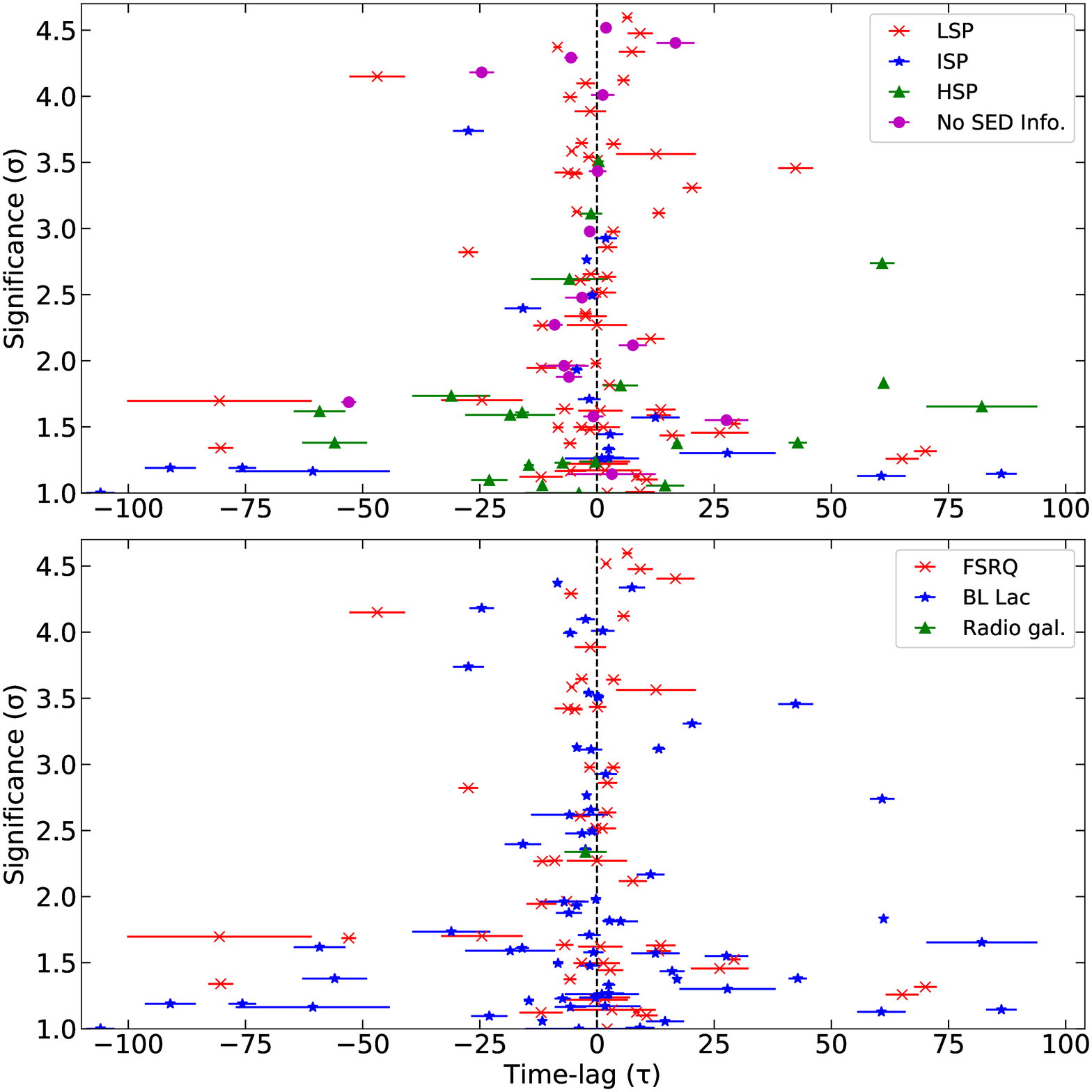} }
 \caption{Time lags versus DCF significance. Top panel: red crosses denote LSPs, blue stars ISPs, green triangles HSPs, 
and magenta circles sources without a synchrotron peak estimate. Bottom panel: red crosses denote FSRQs, blue stars BL Lacs, and green triangles radio galaxies. For positive values the optical emission leads 
the $\gamma$-rays.}
\label{plt:og_lag}
\end{figure}

\begin{figure}
\resizebox{\hsize}{!}{\includegraphics[scale=1]{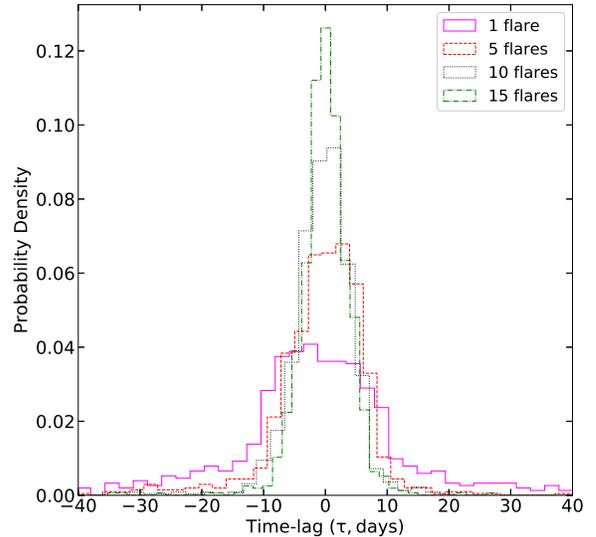} }
 \caption{Distribution of simulated time lags for light curves with 1 (solid pink), 5 (dashed orange), 
10 (dotted black), and 15 simultaneous flares (dash-dotted green). The flares have varying fall times.}
\label{plt:sim_tau}
\end{figure}

\begin{deluxetable*}{ccccccccccccc}
\tablenum{1}\label{tab:sample_properties}
\tablecaption{DCF and flux-flux correlation results for the sources in our sample.}
\tablehead{\colhead{Name} & \colhead{Class}  & \colhead{SED} &\colhead{$z$} & \colhead{$\tau$ (days)} & \colhead{$\sigma_\tau$} & \colhead{DCF Sig.} & \colhead{$m$} & \colhead{$\sigma_m$} & \colhead{$\rho$} & \colhead{$p$-value} & \colhead{Prob. ($m=1$)} & \colhead{Prob. ($m=2$)} }
\startdata
J0017-0512 & F & - & 0.227 & 7.65 & 2.99 & 2.12 & 0.78 & 0.13 & 0.48 & 1.7$\times10^{-07}$ & 0.08 & 0.74 \\ 
J0033-1921 & B & HSP & 0.61 & -22.98 & 3.88 & 1.1 & 0.37 & 0.14 & 0.24 & 0.01 & 0.32 & 0.04 \\ 
J0035+1515 & B & HSP & 1.409 & - & - & - & - & - & - & - & - & - \\ 
J0035+5950 & B & HSP & 0.086 & - & - & - & - & - & - & - & - & - \\ 
J0045+2127 & B & HSP & - & - & - & - & - & - & - & - & - & - \\ 
J0050-0929 & B & ISP & 0.635 & -1.66 & 2.43 & 1.71 & 0.5 & 0.07 & 0.43 & 4.9$\times10^{-09}$ & 0.62 & 0.7 \\ 
J0102+5824 & F & LSP & 0.644 & 12.58 & 8.49 & $>3.5$ & 0.63 & 0.05 & 0.70 & 1.9$\times10^{-21}$ & 0.51 & 0.4 \\ 
J0112+2244 & B & LSP & 0.265 & -1.31 & 1.74 & 2.66 & 1.32 & 0.16 & 0.69 & 7.7$\times10^{-23}$ & 0.01 & 0.08 \\ 
J0115+2519 & B & HSP & 0.37 & -0.35 & 3.49 & 1.24 & 0.44 & 0.2 & 0.31 & 0.0 & 0.26 & 0.56 \\ 
J0132-1654 & F & LSP & 1.02 & -24.57 & 8.71 & 1.7 & 0.49 & 0.12 & 0.34 & 0.0 & 0.97 & 0.47 \\ 
\enddata
\tablecomments{Names are as listed in the KAIT database. The optical classification (Column 2) is taken from Simbad and NED, while the SED classification (Column 3) is from \citealp{Nieppola2006,Nieppola2008,Abdo2010-II,Cohen2014,Lister2015,Mingaliev2015,Angelakis2016} and references therein. Column 4 lists the redshift ($z$). Columns 5, 6, and 7 show the derived time lag, its uncertainty, and the DCF peak significance, respectively. Columns 8 and 9 show the slope ($m$) of the optical--$\gamma$-ray log--log correlation and its uncertainty. Column ``$\rho$'' denotes the Spearman correlation coefficient and column ``$p$-value'' the probability associated with the Spearman test. The last two columns show the probability of association with $m=1$ and $m=2$. The table lists only the first 10 sources. The table is published in its entirety in machine-readable format. A portion is shown here for guidance regarding its form and content.}
\end{deluxetable*}

To study simultaneous interband flux variations, we use the discrete correlation function 
(DCF; \citealp{Edelson1988}) to measure the offset between the optical and $\gamma$-ray light 
curves for all the sources in our sample. Here we provide a brief description of the analysis. A more detailed description is given by \cite{Liodakis2018}. We explore time lags 
in the range [$-600$,600] days with a binning equal to the average sampling of the optical light curves. 
Peaks in the resulting DCFs are fit with a Gaussian function to determine the time lag ($\tau$) 
and its uncertainty ($\sigma_{\tau}$). To estimate the statistical significance of the DCF peaks 
we use the randomization method described by \cite{Cohen2014} and \cite{Liodakis2018}. In summary, we cross-correlate each $\gamma$-ray light curve with optical light curves from all available sources within 3~hr in right ascension from the target source. This ensures similar seasonal optical coverage. With this process we create $\sim7500$ false pairs (depending on the right ascension of the target source) in each time-lag bin. For convenience, we plot the random error distribution as $1\sigma$ (68\%), $2\sigma$ (95\%), and $3\sigma$ (99.7\%) confidence intervals for each source (Fig \ref{plt:multi}).
Since our false-trials distribution covers fluctuations to the $3.5\sigma$ level (expected frequency for normally distributed data 1 in 2149), sources with
a significance above this level are marked as ``$>3.5\sigma$'' in Table \ref{tab:sample_properties},
although we use the actual probabilities estimated for these sources in Fig. \ref{plt:og_lag} and \ref{plt:slope_sigma}.
Figure \ref{plt:multi} shows an example of the light curves (top two rows) and DCF results (bottom-left plot) for J1800+7828. In measuring time
lags for Table \ref{tab:sample_properties}, we use the DCFs computed from the 3~d LAT light curves,
since these probe the smallest $\tau$. The 10~d and 30~d light curves yielded
consistent results within the (larger) errors.

Figure \ref{plt:og_lag} shows the optical to $\gamma$-ray time lags versus 
DCF peak significance. Optical leads the $\gamma$-rays for positive time lags. Of the 178 sources 
in our sample, 121 ($\sim$68\%) show a $>1\sigma$ DCF correlation, while 44 ($\sim$25\%) show a $>2.5\sigma$ correlation. About 74\% of those show a $<3\sigma$ correlation. This is due to sources showing both correlated and uncorrelated events that impact the significance of the DCF. However, $<3\sigma$ sources (although not individually very significant) can still statistically reveal trends between different populations. Not surprisingly, 
time lags close to zero dominate, with half of the sources within 3$\sigma_{\tau}$ of $\tau=0$. 
The large $\tau$ scatter for sources with $<2\sigma$ significance is expected since a number of 
these $\tau$ could be random fluctuations of the DCF and hence randomly distributed. In the full set, we estimate that 11/121 sources represent 
false correlations. For the sources with $>2\sigma$ significant DCF peak, the number of false 
positives drops to 0.31/121 sources. We note that fourteen sources have small $\tau=[-10,10]$ day
lags that are apparently significantly different from zero ($>3\sigma_\tau$). It is possible that the DCF fitting,
which delivers a median $\sigma_\tau=2.5$~d (less than the minimum binning of the DCF), underestimates
the uncertainty. But, if real, the small time lags might reveal differences in the relativistic 
beaming between optical and $\gamma$-rays (e.g., \citealp{Raiteri2011}) or, for EC models, 
differences in the profile of the magnetic and external photon field energy densities with 
distance from the black hole (e.g., \citealp{Janiak2012}). For the high-significance sources it is unlikely that these $\tau$ are due to random fluctuations. Detailed follow-up studies of these individual sources might probe 
the physical mechanisms, but are beyond the scope of the present statistical study. 

	A possible source of apparent lags would be differing flare shapes in the two bands. Since the DCF is sensitive to the centroid of the emission, differing flare shapes could bias the DCF peak position toward either positive or negative values, inducing an apparent lag. To probe the level of this effect, we create simulated light curves for both optical and $\gamma$-rays 
using the {\it astroML} Python package \citep{astroML} to generate a stochastic background with a damped random-walk process and superimposing flares with exponential rise and 
decay profiles. All flares in both bands have matched peak positions, amplitudes, and rise times, 
but the fall times vary randomly between 0.1 and 10 rise times, so that the flares are 
asymmetric. We create light curves of length roughly equal to that observed and sample both bands at 3~d, reproducing the optical observing gaps using observed light curves as templates.  We perform the DCF analysis on these simulated light curves.  

     Figure \ref{plt:sim_tau} 
shows the $\tau$ distributions over $10^4$ simulations, for increasing number of flares. 
While for a single flare the standard deviation of the $\tau$ distribution is about 14~d, for the 15-flare simulation it is $\sim 4$~d. Since we have a mean of 24 flares per source in $\gamma$-rays in the 3~d light curves (and more in the optical; see Sec. \ref{sec:mult_flares}), we
conclude that the effect of random flare-shape changes is less than our typical $\sigma_\tau$.
Of course, if there was a {\it systematic} shape difference between the two bands, this could offset the
apparent flare times. However, even in the extreme case with all fall times a factor of 10 longer than
the rise in one band (in our simulations, rise time is set to 10~d, thus the fall time is 100~d), the resulting induced $\tau$ is $<10$~d. Such strongly asymmetric flares are not seen in the data, and hence we conclude that this effect does not dominate the observed $\tau$.

	Since there seem to be small, but real, $\tau$ in our sample, we attempted to see if there are differences between source classes.  Separating our sample by the synchrotron peak frequency (Fig. \ref{plt:og_lag}, top panel), we find that 
about 80\% of LSPs, 70\% of ISPs, 60\% of HSPs, and 50\% of sources without a SED peak estimate showed a 
$>1\sigma$ correlation. However, there is no difference between the populations' $\tau$ distributions according 
to the Kolmogorov-Smirnov test (K-S test, $p$-value $>0.05$ in all cases).\footnote{The K-S test yields 
the probability of two samples being drawn from the same distribution. For $p$-values $>0.05$ we cannot 
reject the null hypothesis.} Similarly, no difference is seen if we separate by optical spectral
properties into BL Lac and FSRQ (Fig. \ref{plt:og_lag}, bottom panel, K-S test $p$-value 0.77). Similar results were found by \cite{Liodakis2018}. For radio galaxies and unclassified sources, only one radio galaxy shows a significant $\tau$, so we cannot probe the behavior of these source classes.

\section{Optical/$\gamma$-ray Flare Correlations}\label{sec:mult_flares}

\begin{figure}
\resizebox{\hsize}{!}{\includegraphics[scale=1]{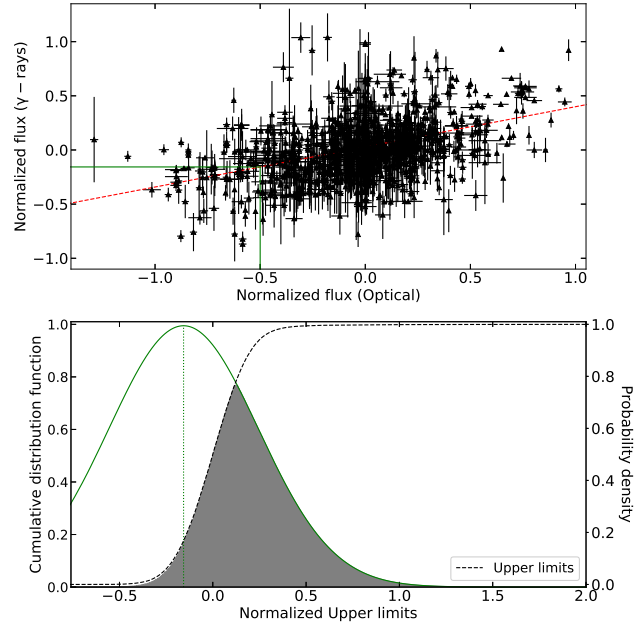} }
 \caption{Orphan flare sensitivity study. Top panel: Logarithm of the normalized optical versus 
$\gamma$-ray flare amplitudes for all of the sources in our sample. The red dashed line shows the 
best-fit relation.  Bottom panel: Distribution of LAT flux detection threshold in LAT light-curve observations,
derived from the cumulative distribution function for the normalized upper limits 
in the $\gamma$-ray light curves. The green solid line contrasts this with an example of the predicted 
$\gamma$-ray flare amplitude for a particular optical flux, given the best-fit flux-flux correlation and RMS scatter.
The gray shaded area thus gives the probability of detecting the $\gamma$-ray flare corresponding 
to the optical event.}
\label{plt:detection_limit}
\end{figure}

\begin{deluxetable*}{lcccc}
\tablenum{2}
\label{tab:numflares}
\tablecaption{Optical and $\gamma$-ray flares association.}
\tabletypesize{\scriptsize}
\tablehead{\colhead{} & \colhead{}  & \colhead{$\gamma$-rays} & \colhead{}  & \colhead{Optical}}
\startdata
& 3~d & 10~d & 30~d & \\
\hline
Total number of flares & 4384 & 2128 & 1017 & 4277\\
Median flare rates & 2.6 & 1.4 & 0.67 & 6.4 \\
Associated flare & 75.3\% & 92.7\% & 93.2\% & 38.0\%/14.0\%/9.0\%\\
Orphan flare & 24.6\% & 14.5\% & 7.7\% & 61.9\%/85.9\%/90.0\%\\
\hline
Median $\tau$ & $-0.76\pm2.5$  & $-0.8\pm 3.9$ & $5.58\pm7.35$ \\
Upper limits & 59.8\% & 29.8\% & 7.7\% &  \\
\enddata
\tablecomments{The flare rates are given in flares per source per year. The uncertainty in the median time lags is the median uncertainty of the time lags. The percentage of associated and orphan events for the optical is for 3~d, 10~d, and 30~d (left to right).}
\end{deluxetable*}

	We wish to characterize the flare activity in the two bands and its correlation in
a model-independent way. We are particularly interested in the incidence of orphan flares --- 
that is, flares appearing in one band, but not the other. These are especially noteworthy events because they could reveal the presence of additional mechanisms and/or serve as signatures of different physical conditions in the jets. To accomplish this, we generate a Bayesian block decomposition of the light curves \citep{Scargle2013}, providing a uniform 
identification of significant flux variations. An example of the Bayesian block analysis is shown 
in Fig. \ref{plt:multi} (second row, right panel). We aligned the blocked light curves using the DCF-determined $\tau$.
``Flares'' are identified as local maxima (center of three Bayesian blocks). Flares in the two bands whose peak blocks overlap are considered associated; those without counterparts in the other band are orphans \citep{Liodakis2018}. 
The optical light curves (nominal cadence $\sim 3$~d, effective cadence typically $<6$~d)
yielded a total of 4277 flares. The number of $\gamma$-ray flares varies with 
binning, with the most found for the 3~d light curves (4443 flares). For the 10~d light curves we find 
2128, and for the 30~d sampling 1017 $\gamma$-ray flares.

	About 10\% of the optical flares occur at the edges of observing seasons, so the peak block
duration cannot be confidently measured. Also, 43\%, 45\%, and 46\%  of the flares in the 3~d, 10~d, and 30~d 
$\gamma$-ray light curves (respectively) occur during optical light curve observing gaps. Considering only flares within overlapping time periods, we find that 
$\sim38.0$\% of optical and 75.3\% of $\gamma$-ray flares are associated in the 3~d light curves (compared to 
14.0\% of optical and 92.7\% of $\gamma$-ray flares in the 10~d light curves, and 9.0\% of optical and 93.2\% of $\gamma$-ray flares in 
the 30~d light curves; Table \ref{tab:numflares}). Thus, decreasing the $\gamma$-ray light curve bin size
has two effects: (1) the fraction of orphan $\gamma$-ray flares increases, and (2) the fraction of orphan 
optical flares decreases. The first effect is due to the narrower 3~d blocks defining local maxima 
better than in the 30~d data; the second occurs since the 3~d blocks resolve more local $\gamma$-ray maxima. Thus, as 
in \citet{Liodakis2018}, we conclude that coarser Bayesian blocks can introduce a number of false flare 
associations. In that work we found that 28\% of $\gamma$-ray flares in 30~d light curves could be falsely associated. In this work, we estimate that no more than 16\%, 20\%, and 30\% of $\gamma$-ray flares could be falsely associated in the 3~d, 10~d, and 30~d light curves, respectively (for details on the false association estimation see \citealp{Liodakis2018}). We can now extrapolate to the limit of infinitely fine $\gamma$-ray sampling by fitting an exponential to the fraction of orphan optical flares versus the binning of the $\gamma$-ray light curves. Interestingly, we find that $\sim50$\% of the optical flares would still lack a $\gamma$-ray counterpart,
implying that there is significant optical activity with little or no $\gamma$-ray response.

Here we next focus on the orphan flares (we discuss the associated flares in Sec. \ref{sec:flare_cor}).	We wish to estimate what fraction of the orphans are simply due to limited sensitivity of their respective observatories.
To this end we fit a linear trend in log--log space to the amplitudes of the well-associated optical ($X$)
and $\gamma$-ray ($Y$) flares, measuring the slope (power-law index) and scatter about this trend. We find the best-fit relation to be $Y = (0.37\pm0.03)X + (0.026\pm0.008)$, while the root-mean-square (RMS) scatter about that line is 0.6. Now, assuming
that the observed trend extends to fainter flares, we can model the expected $\gamma$-ray flux distribution
for any optical flare. This can be compared with the LAT sensitivity distribution to infer the probability
of the LAT detecting the counterpart of that optical flare (Fig. \ref{plt:detection_limit}).
Summing these probabilities, we get the expected total number of LAT detections of optical events at infinite 
$\gamma$-ray sensitivity
(assuming the observed bright-flare trend). We actually observe fewer, indicating true orphans --- that is,
optical flares exhibiting less $\gamma$-flux than the bright-source trend. A similar
exercise can be made for each $\gamma$-ray flare, using the optical observation magnitude uncertainties to
collect the distribution of detection limits. In summary, we find that (for 100\% associations) 2584 optical and 1515 $\gamma$-ray flares could have been associated when only 1174 optical and 1221 $\gamma$-ray flares actually were. This exercise provides our best, sensitivity-corrected estimate of the incidence of orphan flares:  54.5\% of the optical and 20\% of the $\gamma$-ray flares are truly orphan events. 

	Orphan optical variations could be of thermal origin (e.g., accretion-disk outbursts). This might be relevant for high-redshift FSRQs whose optical spectra could be dominated by accretion-disk emission (e.g., \citealp{Paliya2015}). Alternatively, stochastic fluctuations of multiple zones in a turbulent jet (e.g., 
\citealp{Marscher2014, Peirson2018}) may create synchrotron intensity fluctuations along Earth's
line of sight, while not significantly affecting the jet seed photon population. One might expect such
events to be lower in amplitude than associated flares.  Indeed, the median amplitude of the associated flares 
is 1.48 mJy vs. 0.99 mJy ($R$ band) for the orphan flares. The Wilcoxon rank-sum test confirms that the associated flares 
have systematically higher amplitudes than these orphan flares ($p \approx 10^{-14}$), implying that this is not a simple sensitivity effect. This trend was also noted by \cite{Liodakis2018}.  

It is harder to explain $\gamma$-ray orphans in the leptonic 
emission scenario. Local sources of external photons could be seeding $\gamma$-ray flares without significantly increasing the optical emission \citep{MacDonald2015}. The directionality of the magnetic field with respect to the observer's line of sight could also produce apparent orphan flares (e.g., \citealp{Joshi2016}).  Of course, if hadronic emission can dominate the $\gamma$-ray band, one might produce
a GeV flare without concomitant variations in the optical (lepton-produced) flux (e.g., \citealp{Zhang2018}). The possible association 
of a neutrino event with a $\gamma$-ray flare in blazar TXS 0506+056 \citep{IceCube2018} suggests that
this hadronic channel can be present. Since our results show that no more than
$\sim 20$\% of the events are $\gamma$-ray orphans, such flares are rare, contributing no more than 
$\sim 0.3$ flares\,yr$^{-1}$\,source$^{-1}$. Nevertheless, such events could be particularly interesting, since they could reveal the presence of more exotic $\gamma$-ray emission processes.

\section{Interband Flux Correlations}\label{sec:flare_cor}

The strong DCF correlations, small $\tau$, and limited number of orphan $\gamma$-ray
flares all point to leptonic processes --- GeV emission generated by the upscattering of photons from a population of relativistic electrons responsible for the flux variations observed in the optical band, at least for the majority of the events. We would like to identify the nature of this
target photon field. In the simplest leptonic picture, flares are driven by outbursts in the relativistic electron
population produced at the shock. Then, if the jet magnetization is constant, IC emission seeded by photons from an external non-synchrotron source should increase proportionally to the particle density, leading
to linear ($m=1$, in log--log flux space) flare scaling.  In contrast, if the synchrotron emission itself dominates both the optical band and the seed photons (SSC), the $\gamma$-rays 
should vary quadratically with the optical, $m=2$ (e.g., \citealp{Bonnoli2011,Larionov2016}).
While $m=2$ is a fairly robust indication of SSC, $m=1$ flares can occur in both regimes,
if more effects are considered.  For example, magnetic-field fluctuations might drive synchrotron 
variations (with constant particle density), leading to linear SSC flares. Still, we can test the simple EC/SSC split by measuring the flux ratios of the correlated variability.

\subsection{Raw Slope Estimation}

\begin{figure}
\resizebox{\hsize}{!}{\includegraphics[scale=1]{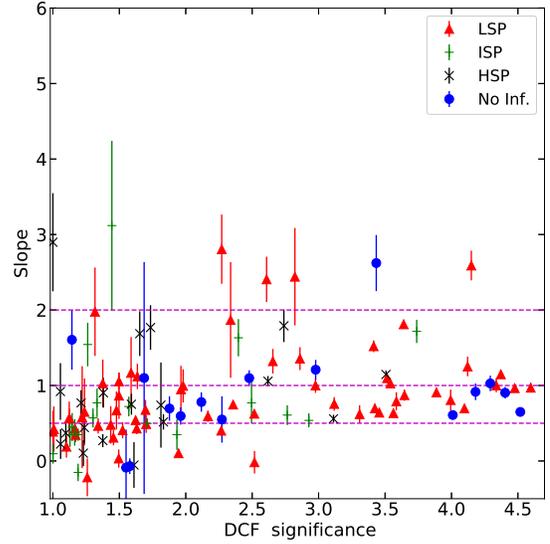} }
 \caption{Slopes $m$ of the optical--$\gamma$-ray flux relation versus the significance of the DCF peak. 
The red triangles show LSPs, green pluses ISPs, black crosses HSPs, and blue circles indicate sources with 
no SED peak information. The dotted lines show $m=0.5$, $m=1$, and $m=2$}
\label{plt:slope_sigma}
\end{figure}

\begin{figure}
\resizebox{\hsize}{!}{\includegraphics[scale=1]{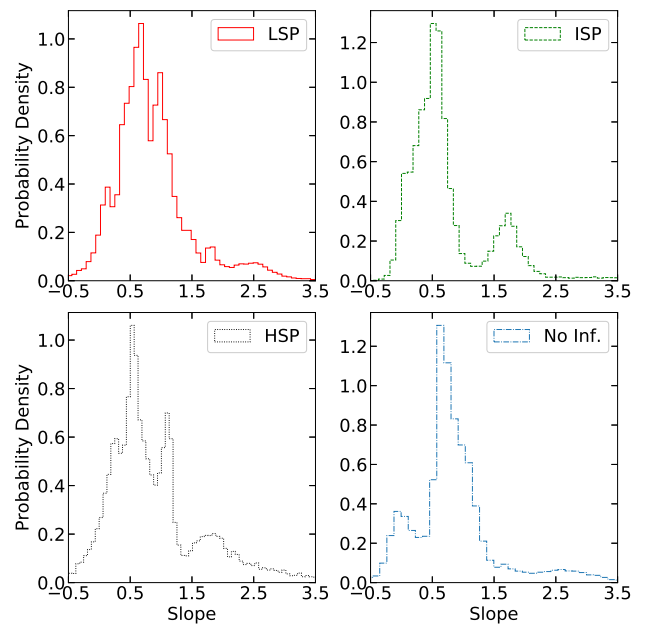} }
 \caption{Posterior distribution of the $m$ for our sample. 
Upper-left panel LSPs, upper-right ISPs, lower-left HSPs, and lower-right sources with 
no SED peak information.}
\label{plt:slope_posterior}
\end{figure}

We start from the light-curve Bayesian block decompositions, aligned by the DCF-determined $\tau$. For this exercise we use the 10~d binned light curves which provide a good compromise 
between time resolution and sensitivity. This decreases the number of upper-limits intervals in the light 
curves (Table \ref{tab:numflares}) while retaining a sampling rate close to the effective $\sim 6$~d cadence 
of the optical light curves. Since the optical light curves have shorter coverage and are more variable, we use the optical flux Bayesian blocks to group the DCF lag corrected $\gamma$-ray observations. We thus compare not only the individual bright flares but also the lower amplitude
variability common to both bands, taking both short and long term fluctuations into account. When all blocks are used, this does allow pollution of the intraband relation by orphan flare intervals. We discuss correction for this effect in Sec. \ref{subsec:rem_orphans} and Sec. \ref{subsec:stat__corr_orphans}.

	For each source, we compare the optical and $\gamma$-ray flux variations using the Spearman rank-order 
correlation test (Spearman test).\footnote{The Spearman test is a nonparametric measure of the monotonic correlation 
between two samples. The correlation coefficient $\rho$ takes values $[-1,1]$, with 1 denoting a perfect positive
correlation, 0 no correlation, and $-1$ a perfect negative correlation.} We also fit a linear model in log--log space 
using the Bayesian regression method described by \cite{Kelly2007}, taking the uncertainties of both the optical 
and $\gamma$-rays into account. The slope $m$ of the linear model probes the dominant mechanism of $\gamma$-ray
production in each source. Figure \ref{plt:slope_sigma} shows the results for the slopes versus the significance 
of the DCF for the sources in our sample. Out of the sources in our sample, 18 do not show a statistically 
significant correlation according to the Spearman test ($p$-value of a correlation arising by chance $>5\%$). 
These are sources where flux variations are dominated by uncorrelated fluctuations. Not surprisingly, most of 
the sources without a significant correlation coefficient lie in the low-significance DCF peak range. 

For all sources with a $>1\sigma$ correlation, we sample a hundred estimates from the posterior distribution of each measured $m$. We plot all those slope distributions together in Fig. \ref{plt:slope_posterior}, showing the range of slopes for a set of sources. Dividing by source class, we see a tail to large $m$ and apparent bimodality (especially for ISP sources) with peaks consistent with the theoretical expectations.

\subsection{Effect of Orphan Flares}	

\begin{figure}
\resizebox{\hsize}{!}{\includegraphics[scale=1]{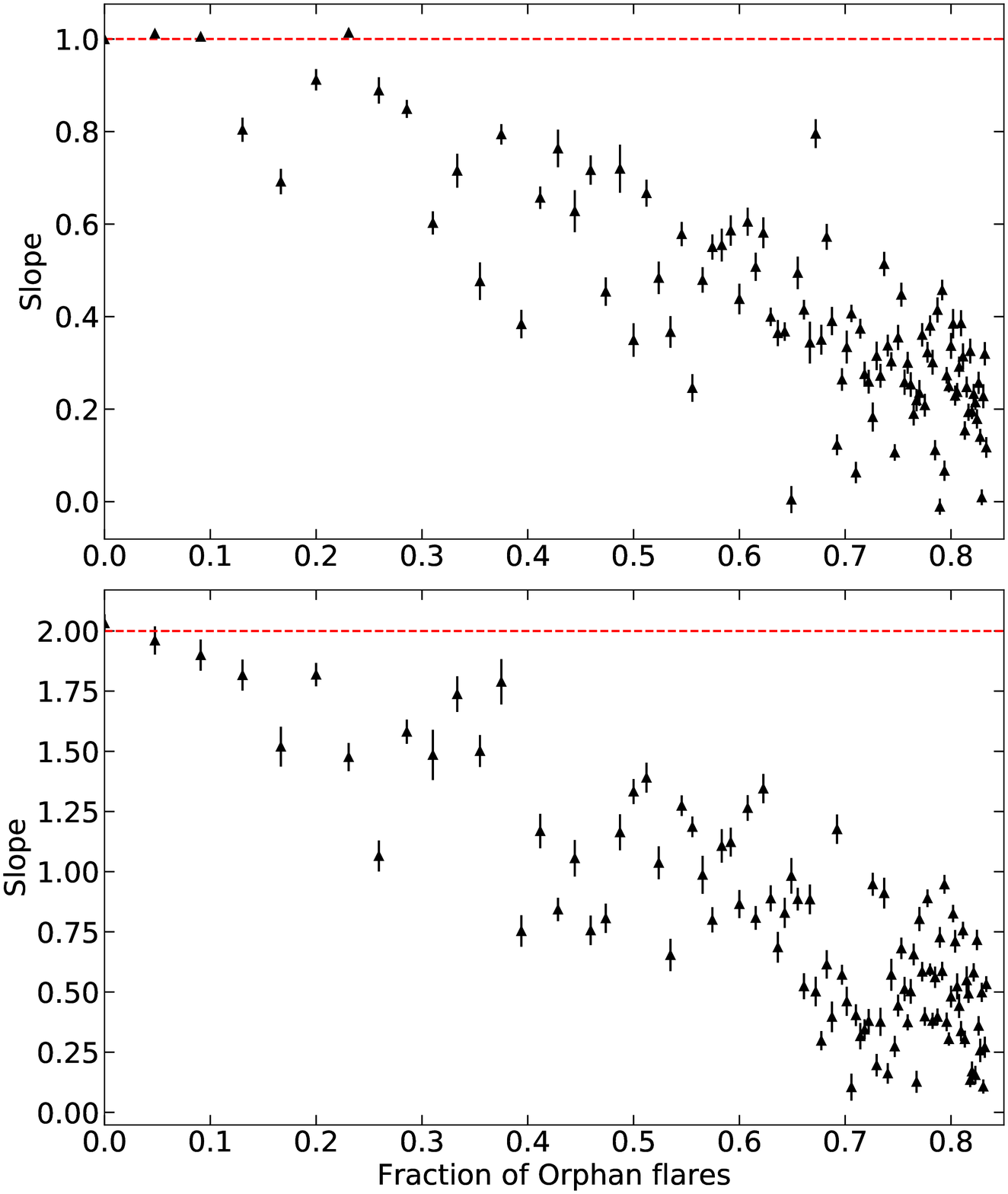} }
 \caption{The effect of orphan optical flares on measured slope $m$. Top panel, $m=1$; Bottom panel, $m=2$}.
\label{plt:sim_orphan_flares}
\end{figure}
	
	While the simple model suggests that flares should follow either $m=1$ or $m=2$, and indeed the population appears to be consistent (within the spread) with this picture (Fig. \ref{plt:slope_posterior}), we see that many individual measurements lie significantly away from these values. We can understand this by considering that any
given source may display more than one type of fluctuation. Certainly values $1<m<2$ can be 
produced by mixtures of comparable numbers of SSC and EC events. Also, if Klein-Nishina cutoffs suppress
the emission in some $\gamma$-ray events \citep{Petropoulou2015}, this can drive the mean slope below the expected value. However, as seen in
the posterior histograms, the most common effective $m$ is $<1$. This is expected, as these light
curves contain a significant number of orphan flares, especially in the optical. By definition, the $\gamma$-ray
flux shows no correlation with optical emission during optical orphans, and so these events add a component of $m \approx 0$ fluctuations. Since this must substantially affect the direct $m$ estimates, we seek to correct
for orphan events before further discussing the optical--$\gamma$-ray correlation.

	We attempt to correct for this effect by running a sequence of simulations. As a first reality
check, we generate uncorrelated light-curve pairs by cross-matching sources and assigning a typical random
$-30{\, \rm d} < \tau < 30{\, \rm d}$ lag, and fit to the Bayesian-block flux-flux distribution. As expected,
the resulting $m$ scatters closely around zero. We then check that light curves with correlated fluctuations
deliver the expected slope, by following the procedure in Sec. \ref{sec:cross} to generate
artificial light curves with simultaneous flares in the two bands. Typically we generate 20 flares per light curve
and fit the fluctuation distributions. Proportional flares lead to $<m>=1$, while quadratically scaled $\gamma$-ray
flares give $<m>=2$. We next add orphan optical events, increasing from 0 to 100 orphan flares per light curve
and measuring the slope in each of many realizations. Figure \ref{plt:sim_orphan_flares} shows the derived $m$
versus the orphan flare fraction, for a true $m=1$ (top panel) and $m=2$ (bottom panel). The clear decrease,
even beyond the observed $m_1 \approx 0.7$ and $m_2 \approx 1.7$, can be seen. Of course, orphan $\gamma$-ray flares
have the opposite effect, increasing $m$ --- but given their low rate, the main effect is from the optical events.

\subsection{Removing Orphan Flares}\label{subsec:rem_orphans}

\begin{figure}
\resizebox{\hsize}{!}{\includegraphics[scale=1]{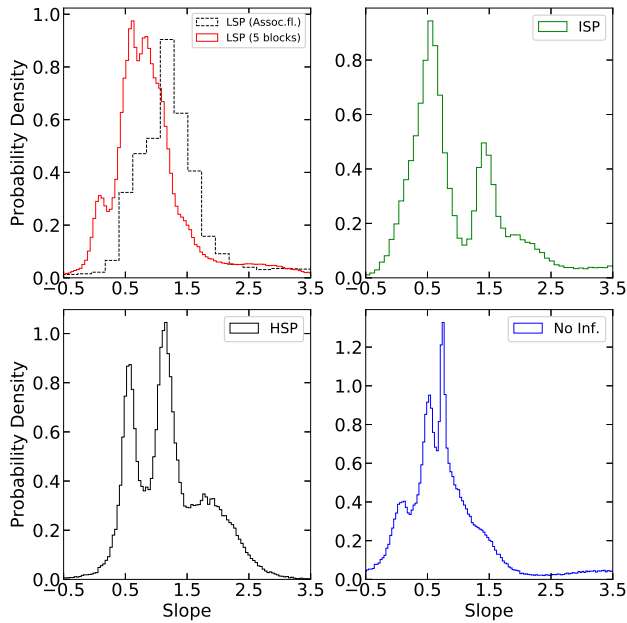} }
 \caption{Posterior distribution of the $m$ for our sample after having removed 5 Bayesian blocks associated with orphan events. Upper-left panel LSPs, upper right ISPs, lower left HSPs, and lower right sources with no SED peak information. In the case of LSPs we additionally plot the posterior $m$ distribution using only the associated flares (dashed black line).}
\label{plt:cut_orphan_flares}
\end{figure}

We can, to some extent, directly measure the relationship excluding the orphan flares.
If we use {\it only} the flare intervals themselves, too few ($\sim 25\%$) of the sources (majority LSPs) have enough 
common flares to measure both the mean flux scaling (effectively correlation intercept) and correlation slope $m$.
Instead, we excise the times associated with the optical blocks of known orphan optical flares in the 
light curves, leaving both the known correlated flares and low-level flux variations that may include weak flaring in the
two bands. This excises the optical orphans strongly affecting $m$, while leaving enough flux-flux measurements
to anchor the intercept and slope measurement for nearly all sources. In practice, we excise 5 blocks
around each orphan optical maximum. 

The resulting $m$ distributions for the four blazar classes
are shown in Figure \ref{plt:cut_orphan_flares}.  Stronger peaks now appear in the distributions near the expected intrinsic slopes, in particular for HSPs, implying that a fraction of these sources are SSC dominated, while others show many EC events. It is interesting that for some sources $m=0.5$ appears to persist even after excising orphan events. Whether there is a physical origin to such $m$ values is unknown (e.g., proton synchrotron could create sublinear relations; \citealp{Mastichiadis2013}), although one likely interpretation, already discussed above, is simply that uncorrelated long-term fluctuations on top of a linear scaling could yield $m=0.5$.

\subsection{Statistical Correction of $m$}\label{subsec:stat__corr_orphans}

\begin{figure}
\resizebox{\hsize}{!}{\includegraphics[scale=1]{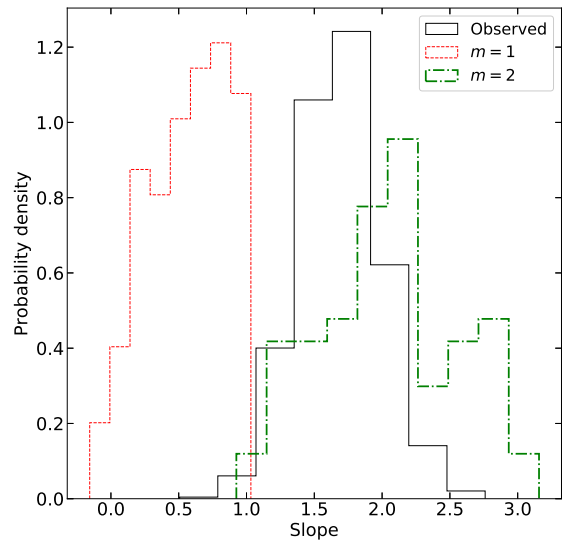} }
\caption{Posterior distribution of the observed $m$ for J0136+3905 compared with the expected distributions for $m=1$ and $m=2$ given the observed orphan flare fraction.}
\label{plt:EC_SSC_observed}
\end{figure}

\begin{figure}
\resizebox{\hsize}{!}{\includegraphics[scale=1]{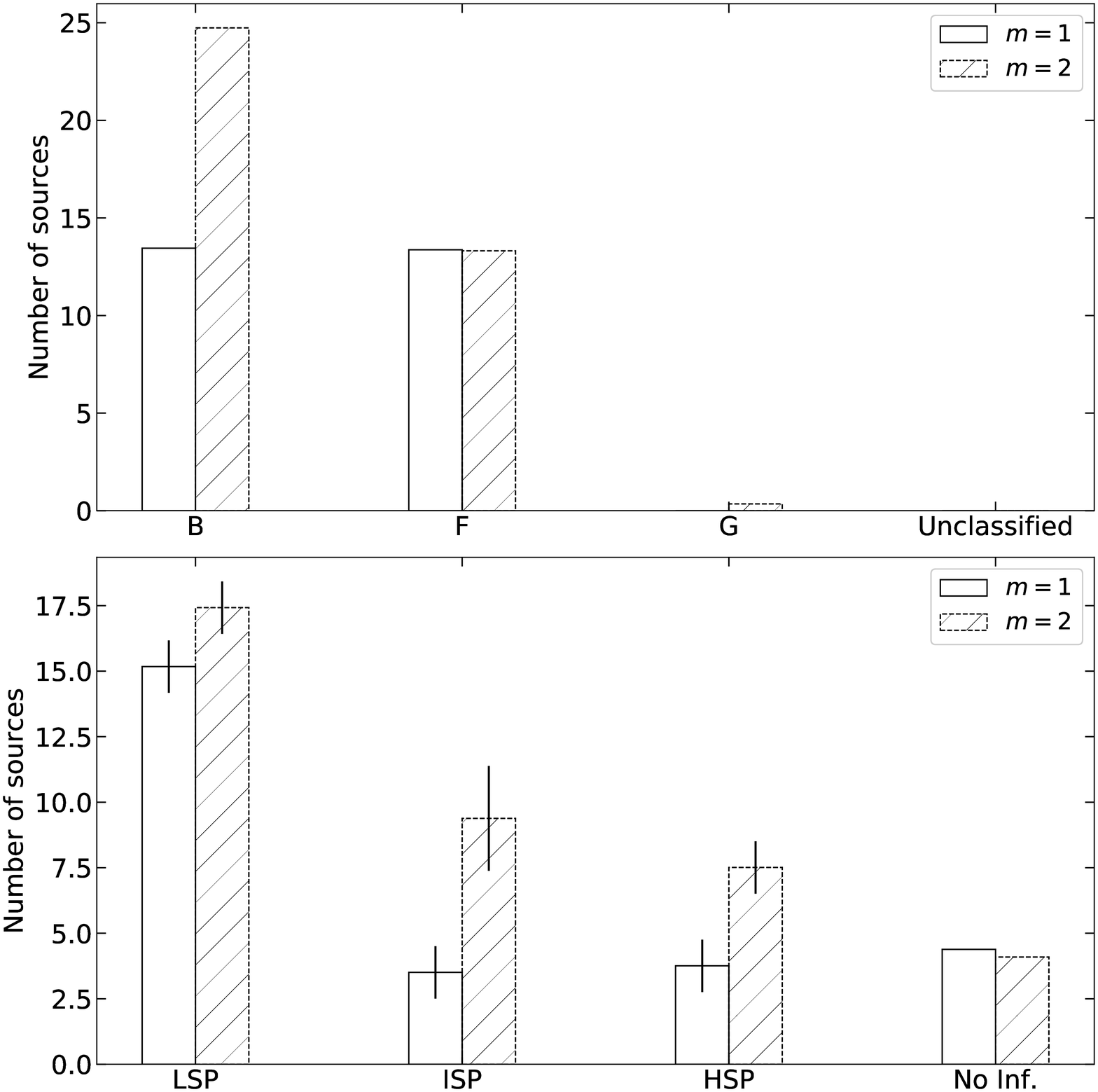} }
\caption{Number of sources best described by $m=1$ (solid) or $m=2$ (dashed) for the different 
populations in our sample, after correction for the affect of the orphan flare fraction observed in each
source. Top panel: separation by optical spectral properties (B is for BL Lacs, F for FSRQs, and G for radio galaxies). Bottom panel: separation by $\nu_{\rm sy}$. The error bars in the lower panel indicate the bin uncertainty attributable to imprecision
in locating the synchrotron peak ($\log(\sigma_{\nu_{\rm sy}})\pm0.3$).}
\label{plt:barplot}
\end{figure}

We can now attempt to statistically correct the measured $m$ values for orphan
flare pollution using simulated light curves. For each source we have a known number of orphan optical and 
$\gamma$-ray flares. If we assume
a starting value $m=1$ or $m=2$, we can simulate a light curve and extract a value of $m$. Repeating this process 100 times allows us to predict
the resulting distribution of measured $m$ for the sources' orphan flare fraction. We then compare the measured $m$ posterior distribution
with these distributions. Their overlap points to whether it is more likely the result of $m=1$ or $m=2$ initial values. Figure \ref{plt:EC_SSC_observed} shows an example of the overlap between observed and expected distributions given an initial value of $m=1$ and $m=2$ and the associated and orphan flare fraction in J0136+3905. In this case, there is a 82\% overlap with the distribution of an initial $m=2$ while only 0.9\% for an initial $m=1$. Summing the probabilities of all the sources yields the expected number of sources in different blazar classes statistically consistent with either $m=1$ or $m=2$ (Fig. \ref{plt:barplot}). 

 About half of the FSRQs favor either a linear (50.1\%) or a quadratic relation, while more than half of the BL Lacs (64.7\%) match a quadratic relation and thus SSC. This is hardly surprising
given the dominance of jet synchrotron emission in the BL Lacs' optical spectra.  More interesting is the 
$\nu_{\rm sy}$ partition in the bottom panel. LSPs show a mild (53.4\%) preference to a quadratic SSC relation, while ISPs and HSPs show a clear preference at 72.7\% and 66.7\% (respectively). Sources without SED peak classification exhibit a mild apparent preference for a linear relation (51\%); however, no conclusion can be made given the unknown composition of this subsample. Since the $\nu_{\rm sy}$ values themselves are imperfectly measured (typical $\log(\sigma_{\nu_{\rm sy}})$ = $\pm 0.3$; \citealp{Lister2015}), we can estimate the resulting variation in our $m=1$ and $m=2$ values; the results are shown by the histogram error bars.

The results for the flux-flux correlation analysis for all the sources in our sample are listed in Table \ref{tab:sample_properties}.

\section{Conclusions}\label{sec:disc-conc}

We have explored the connection between optical and $\gamma$-ray light-curve flares for a large number of blazars,
studying correlations both in flare time and flare amplitude. We found 121 sources with measured ($>1\sigma$) 
DCF correlations (44 sources show $>2.5\sigma$ correlations); the majority of their derived time lags are close to 0 (within 20\,d).
This strongly favors a common origin for flares in the two bands, implying that leptonic processes dominate 
blazar $\gamma$-rays without excluding the possibility of hadronic contribution to the emission. Although most time lags are small (in the range $[-10,10]$\,d), some (12 sources) exhibit lags significantly different from zero. We found it
unlikely that varying flare shapes could account for this. The most significant lag sources should be
investigated in detail to probe the origin of this effect. 

With a Bayesian block \citep{Scargle2013} decomposition of the light curves we obtained a uniform selection 
of flares for each source in both wavebands, which were matched using the measured time lags $\tau$.
Modeling the interband flux correlation and its distribution, we were able to correct 
for the limitations of $\gamma$-ray time resolution and observational sensitivity, ascertaining for
the first time the true rates of optical and $\gamma$-ray orphan flares. We find that 54.5\% of the optical and
20\% of the $\gamma$-ray flares are true orphan events. The optical orphans are generally lower level
fluctuations (perhaps disk or apparent fluctuations of a multi-zone jet). The rarer $\gamma$-ray orphans are more puzzling,
but might be induced by differences in local beaming or by hadron-generated outbursts.

The Bayesian block light curves also let us compare the correlation in flare fluxes. We started from the hypothesis (usually favored in one-zone leptonic models) that EC events induce linear ($m=1$) fluctuations, while SSC events generated by 
electron number increases will have a quadratic ($m=2$) $\gamma$-ray response. We found that sources 
can be dominated by either event class, but that the presence of the uncorrelated optical orphans 
systematically decreases the effective $m$ measured in the actual light curves. A simple
excision of the orphan flare intervals appears to improve measurement of $m$.  By simulating this effect, we developed a method to correct for the known fraction of orphan events in individual sources, to
recover the most likely intrinsic value of $m$. While sources still display a variety of behaviors, perhaps
allowing both EC and SSC flares in a given source, there are still interesting trends in the populations.
For example, higher peaked ISPs and HSPs are more likely to display $m=2$ (SSC-like) correlations. Certain hadronic models can also produce correlated variability with intraband scalings similar to those found in this work \citep{Mastichiadis2013}. It would be interesting to compare the predictions of such models for individual sources to the observed slopes (Table \ref{tab:sample_properties}).

	We find that a statistical treatment of correlated flare variability in a large number
of blazars has both clarified and complicated the interpretation of these events. LSPs reveal only a mild preference for quadratic (SSC-like) fluctuations, while a much clearer trend for SSC processes 
is seen for many ISPs and HSPs; however, the separation into source classes is far from pure, and all classes
show sources dominated by events from either type. Indeed, while lepton-dominated IC emission 
appears to dominate the population, there is still room for a (demonstrably small) admixture of
hadron-generated events in the orphan $\gamma$-ray flare sample. 

\acknowledgements

The authors thank Lea Marcotulli, Alberto Dominguez, Vaidehi Paliya, Jeremy Perkins, and the anonymous referee for comments and suggestions that helped improve this work. The \textit{Fermi} LAT Collaboration acknowledges generous ongoing support from a number of agencies and institutes that have supported both the development and the operation of the LAT as well as scientific data analysis.  These include NASA and the Department of Energy in the United States, the Commissariat \`a l'Energie Atomique and the Centre National de la Recherche Scientifique/Institut National de Physique Nucl\'eaire et de Physique des Particules in France, the Agenzia Spaziale Italiana and the Istituto Nazionale di Fisica Nucleare in Italy, the Ministry of Education, Culture, Sports, Science, and Technology (MEXT), High Energy Accelerator Research Organization (KEK), and Japan Aerospace Exploration Agency (JAXA) in Japan, and the K.~A. Wallenberg Foundation, the Swedish Research Council, and the Swedish National Space Board in Sweden. Additional support for science analysis during the operations phase is gratefully acknowledged from the Istituto Nazionale di Astrofisica in Italy and the Centre National d'\'Etudes Spatiales in France. This work performed in part under DOE Contract DE-AC02-76SF00515.

This work was financed in part by NASA grants NNX10AU09G, GO-31089, NNX12AF12G, and NAS5-00147, as well as by {\it Fermi} Guest Investigator grants NNX08AW56G, NNX09AU10G, NNX12AO93G, and NNX15AU81G. A.V.F. and W.Z. are also grateful for support from the Christopher R. Redlich Fund, the TABASGO Foundation, National Science Foundation (NSF) grant AST-1211916, and the Miller Institute for Basic Research in Science (U.C. Berkeley). This research has made use of data from the robotic 0.76-m Katzman Automatic Imaging telescope at Lick Observatory. We thank the late Weidong Li for setting up the KAIT blazar monitoring program. KAIT and its ongoing operation were made possible by donations from Sun Microsystems, Inc., the Hewlett-Packard Company, AutoScope Corporation, Lick Observatory, the NSF, the University of California, the Sylvia and Jim Katzman Foundation, and the TABASGO Foundation. Research at Lick Observatory is partially supported by a generous gift from Google.   This paper has made use of up-to-date SMARTS optical/near-infrared light curves that are available at www.astro.yale.edu/smarts/glast/home.php. Data from the Steward Observatory spectropolarimetric monitoring project were used. This research made use of the SIMBAD database, operated at CDS, Strasbourg, France \citep{Wenger2000}; it also made use of the NASA/IPAC Extragalactic Database (NED), which is operated by the Jet Propulsion Laboratory, California Institute of Technology, under contract with NASA.

\facilities{{\it Fermi}, Lick/KAIT, SMARTS, Steward Observatory}

\bibliographystyle{aasjournal}
\bibliography{bibliography} 
\end{document}